# *In Situ* Phase-Transition Crystallization of All-Inorganic Water-Resistant Exciton-Radiative Heteroepitaxial CsPbBr$_3$–CsPb$_2$Br$_5$ Core–Shell Perovskite Nanocrystals

Tianyuan Liang, Wenjie Liu, Xiaoyu Liu, Yuanyuan Li, Wenhui Wu, and Jiyang Fan*

School of Physics, Southeast University, Nanjing 211189, P. R. China

**Corresponding Author**

*E-mail: jyfan@seu.edu.cn

**ABSTRACT**: The instability of metal halide perovskites upon exposure to moisture or heat strongly hampers their applications in optoelectronic devices. Here, we report the large-yield synthesis of highly water-resistant total-inorganic green luminescent CsPbBr$_3$/CsPb$_2$Br$_5$ core/shell heteronanocrystals (HNCs) by developing an in situ phase transition approach. It is implemented via water-driven phase transition of the original monoclinic CsPbBr$_3$ nanocrystal and the resultant tetragonal CsPb$_2$Br$_5$ nanoshell has small lattice mismatch with the CsPbBr$_3$ core, which ensures formation of an epitaxial interface for the yielded CsPbBr$_3$/CsPb$_2$Br$_5$ HNC. These HNCs maintain nearly 100% of the original luminescence intensity after immersion in water for eleven months and the luminescence intensity drops only to 81.3% at 100 °C. The transient luminescence spectroscopy and density functional theory calculation reveal that there are double radiative recombination channels in the core CsPbBr$_3$ nanocrystal, and the electron potential barrier provided by the CsPb$_2$Br$_5$ nanoshell significantly improves the exciton recombination rate. A prototype quasi-white light-emitting device based on these robust CsPbBr$_3$/CsPb$_2$Br$_5$ HNCs is realized, showing their strong competence in solid-state lighting and wide color-gamut displays.

## 1. INTRODUCTION

In the past few years, the metal halide perovskites have attracted great interest due to their very wide crystal structure spectrum and excellent electrical and optical properties.[1,2] For example, the power conversion efficiency of the perovskite solar cells has reached up to 25% and is comparable to that of the conventional silicon solar cells.[3] Although there could be the quantum efficiency droop (gradual drop) problem,[4] the investigators have tried to fabricate perovskite light-emitting devices with external quantum efficiencies > 20%.[5–7] Some novel types of perovskites such as layered perovskites[8] and zero-dimensional perovskites[9] show broadband white luminescence[10] or near-UV luminescence [such as self-trapped exciton quantum confinement luminescence of Cs$_4$PbCl$_6$ or Cs$_4$PbBr$_6$[11,12]], and they bring more hopes for realization of perovskite-based optoelectronic devices. Although perovskites have many virtues, but their instability upon exposure to moisture or heat strongly hampers their practical applications.[13–15] This is particularly serious for the self-assembly crystallized organic–inorganic hybrid perovskites[16] and limits their device applications.[17–19] Another large family of perovskites are the all-inorganic CsPbX$_3$ (X = Cl, Br, and I)[20] and their nanocrystals (NCs) have aroused high interest owing to better stability and good luminescence properties.[21–24] However, even the CsPbX$_3$ NCs have poor stability upon exposure to humidity, light, or heat.[25–28] The studies of the conventional core/shell semiconductor heterostructures reveal that the existence of a shell can significantly influence the migration of the photogenerated carriers in the core and control the luminescence properties.[29–34] The shell also improves stability of the core nanocrystal.[35,36] Therefore, some researchers have conceived to synthesize core–shell nanostructures to improve stability of the perovskite NCs.[37–39] However, because the perovskites have fragile ionic structures, the previous study has been limited to synthesis of the core–shell perovskite nanostructures employing various kinds of amorphous oxides ($SiO_2$, $TiO_2$, and $ZrO_2$) as the encapsulation layers.[40–43] There have also been reports on synthesis of the composite materials composed of perovskites and chalcogenides (CdS, ZnS, PbS, and PbSe).[44–47] For most of these materials, it is unclear whether the perovskite is epitaxially fused with another material.[37]

The previous study revealed that the synthesized millimeter-sized $CsPbBr_3$ crystals contained a small amount of $CsPb_2Br_5$ (10%) as the byproduct.[48] In fact, the ternary phase diagram of Cs, Pb, and Br (Figure 1b) indicates that monoclinic $CsPbBr_3$, tetragonal $CsPb_2Br_5$ (Figure 1c), and hexagonal $Cs_4PbBr_6$ lie on the same segment whose endpoints are $PbBr_2$ and CsBr. This explains why $CsPbBr_3$ and $CsPb_2Br_5$ often coexist in the product, although they are only a mixture[49–51] rather than heteronanocrystals (HNCs). There have been reports on synthesis of dual-phase $CsPbBr_3$–$CsPb_2Br_5$ perovskite films and dual-phase halide perovskite nanorings, which are not core–shell structures.[52,53] On the other hand, it has been known that $CsPbBr_3$ may decompose into $CsPb_2Br_5$ in water following $2CsPbBr_3 \xrightarrow{H_2O} CsBr + CsPb_2Br_5$.[54] Herein, we develop an in situ phase transition strategy to synthesize highly water-resistant luminescent $CsPbBr_3$/$CsPb_2Br_5$ core/shell HNCs. They have good epitaxial crystallization at the core–shell interface. Both their microstructures and luminescence properties exhibit high water and thermal stability. The implementation of a prototype white light-emitting diode (LED) based on these HNCs demonstrates their strong application potential in lighting and displays.

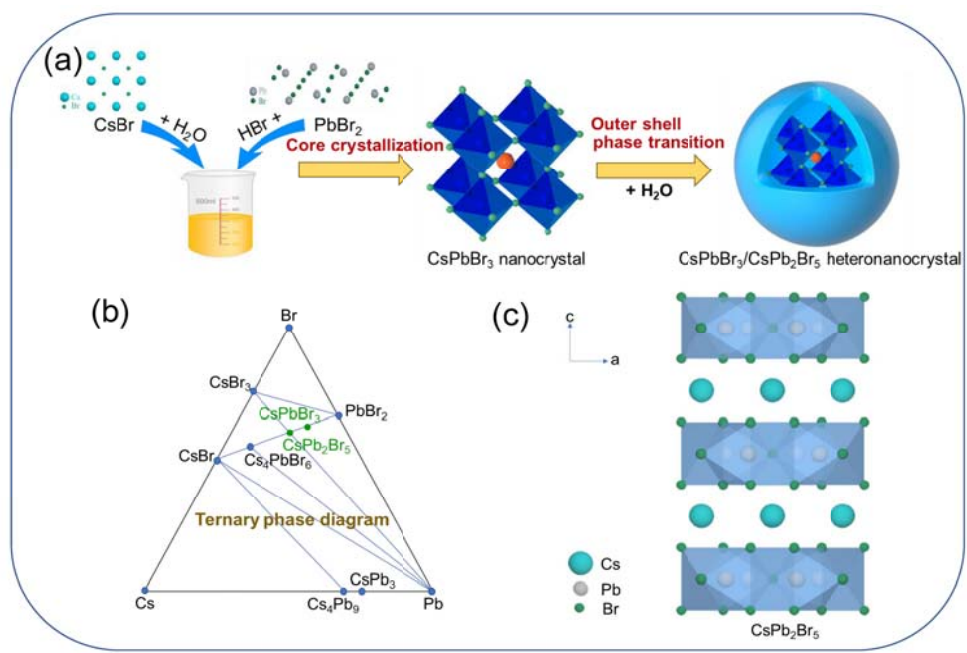

**Figure 1.** (a) Schematic diagram showing synthesis strategy of the $CsPbBr_3$/$CsPb_2Br_5$ core/shell HNC. (b) Ternary phase diagram of Cs, Pb, and Br. $CsPbBr_3$ and $CsPb_2Br_5$ lie between the precursors CsBr and $PbBr_2$ on the segment. (c) Crystal structure of tetragonal $CsPb_2Br_5$.

## 2. RESULTS AND DISCUSSION

**2.1. Microstructure of Phase Transition Induced Heteronanocrystals.** Figure 1a shows the synthesis strategy of the $CsPbBr_3$/$CsPb_2Br_5$ core/shell HNCs. In brief, $PbBr_2$ in hydrobromic acid and CsBr in deionized water reacted to yield an orange powder containing $CsPbBr_3$ NCs. It should be noted that the $CsPbBr_3$ NCs (orange powder) were synthesized in the hydrobromic acid solution and only a small amount of deionized water [CsBr (10 mmol) in water (3 mL) mixed with $PbBr_2$ (10 mmol) in HBr (8 mL)] was used to dissolve CsBr. When the synthesized $CsPbBr_3$ orange powder was immersed in a large amount of deionized water and stirred, the surfaces of the $CsPbBr_3$ NCs lost CsBr owing to the action of water, leading to their in situ phase transition to become $CsPb_2Br_5$. The generated $CsPb_2Br_5$ on the surfaces of the $CsPbBr_3$ NCs prevented water from further penetrating into the interior of the $CsPbBr_3$ NCs and thus the $CsPbBr_3/CsPb_2Br_5$ core/shell HNCs (white powder) formed. The orange powder exhibited no naked-eye-identifiable photoluminescence (PL) under UV-light excitation (inset of Figure 2a). Figure 2a shows its X-ray diffraction (XRD) pattern, and most of the diffraction peaks can be identified as those of monoclinic $CsPbBr_3$ (PDF#18-0364), but some weaker peaks are identified as those of tetragonal $CsPb_2Br_5$ (PDF#25-0211). This agrees with the previous finding that the synthesized $CsPbBr_3$ crystallites often contain a small amount of $CsPb_2Br_5$.[48] The detailed comparison (Figure S1) confirms that the orange powder is indeed monoclinic $CsPbBr_3$ rather than the other structured $CsPbBr_3$. Unlike monoclinic $CsPbBr_3$ that is composed of interconnected $[PbBr_6]^{4-}$ octahedra (Figure 2c), tetragonal $CsPb_2Br_5$ is composed of quasi-two dimensional periodic $[Pb_2Br_5]^-$ layers that are separated by the intermediate $Cs^+$ layers (Figures 1c and 2c). The white powder exhibits very bright green luminescence under UV-light excitation (inset of Figure 2b). Its XRD pattern (Figure 2b) contains mainly diffraction signals of tetragonal $CsPb_2Br_5$, but it also contains some weak diffraction peaks belonging to monoclinic $CsPbBr_3$. Since $CsPbBr_3$ decomposes readily in water, so it is strange why the white powder transformed from the original orange powder in water contains $CsPbBr_3$. The transmission electron microscopy (TEM) reveals the reason. The high-resolution TEM (HRTEM) image of the orange powder (Figure S2a) reveals that the orange powder contains agglomerates composed of closely and tightly bound nanoparticles with sizes of 10–36 nm (Figure S3). These nanoparticles have good crystallinity (Figure S2b–d), and the measured lattice spacing values are 4.13 and 2.92/2.91

Å, being consistent with that of the (110) and ($\bar{2}$00) planes of monoclinic $CsPbBr_3$. Figure S2e shows the corresponding selected-area electron diffraction (SAED) pattern, and the bright diffraction spots are separately ascribed to the (110), ($\bar{2}$00), and ($\bar{2}$20) planes of monoclinic $CsPbBr_3$. The $CsPb_2Br_5$ nanoparticles are not identified from the HRTEM images, suggesting their quantity should be small and thus can only be identified from the XRD pattern.

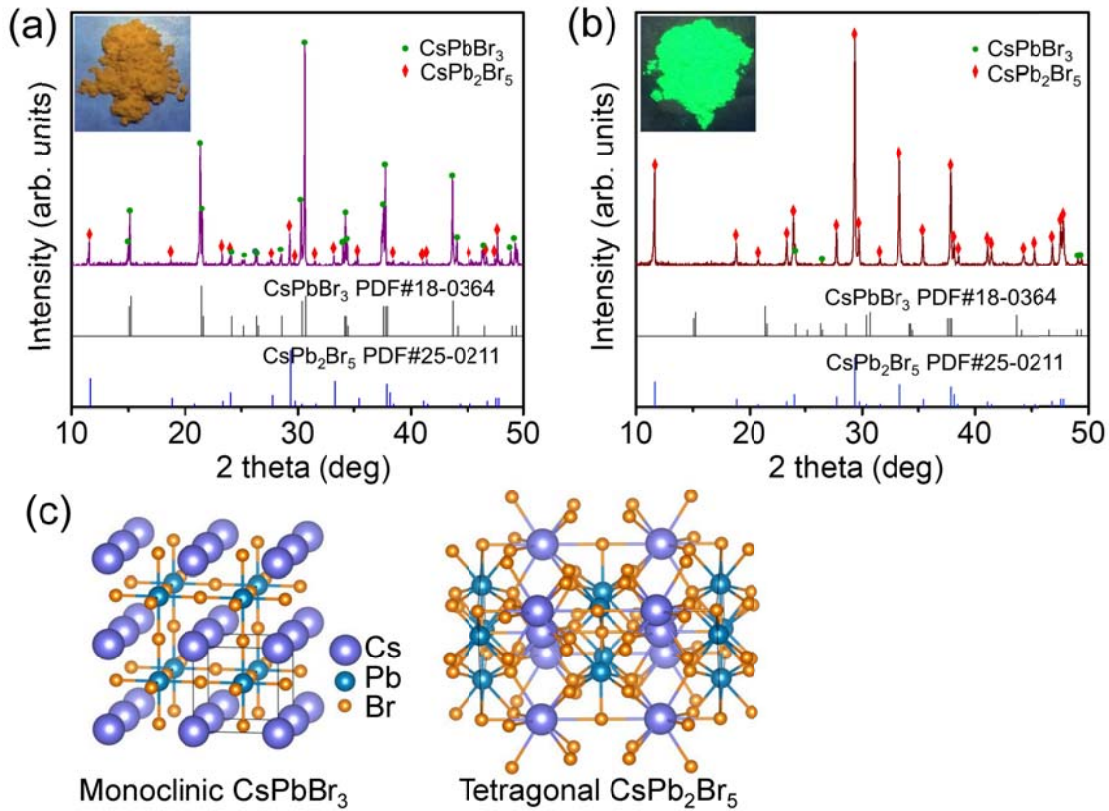

**Figure 2.** XRD patterns of (a) orange powder and (b) white powder. Blue bars show standard XRD patterns of monoclinic $CsPbBr_3$ and tetragonal $CsPb_2Br_5$. Inset shows luminescence photos under 365-nm excitation. (c) Crystal structures of monoclinic $CsPbBr_3$ and tetragonal $CsPb_2Br_5$.

The HRTEM image of the white powder (Figure 3a) reveals that it is composed of individual core/shell NCs with sizes of 15–33 nm and core sizes of 4–14 nm (Figure S4). The ratios of the diameters of the cores and core/shell structures range from 0.2 to 0.5. The contrast of the core and shell exhibited in the TEM image is remarkable, suggesting their different compositions. Figure 3b shows the fast Fourier transform (FFT) of the HRTEM image in Figure 3a, and bright spots correspond to the (310) planes of $CsPb_2Br_5$ and ($\bar{2}$00) and ($\bar{1}$11) planes of $CsPbBr_3$, respectively (Figure 3c). Figure 3d shows a magnified HRTEM image of a typical core/shell nanoparticle. The lattice spacing of the core is 2.90 Å, being consistent with that of the ($\bar{2}$00) planes of monoclinic $CsPbBr_3$, and the lattice spacing of the shell is 4.23 Å, corresponding to the (200) planes of tetragonal $CsPb_2Br_5$. Figure 3e shows the FFT of Figure 3d. It comprises bright spots belonging to the (200) planes of $CsPb_2Br_5$ and ($\bar{2}$00) planes of $CsPbBr_3$ (Figure 3f), and the angle between these planes is derived to be 45°. Figure 3g shows the corresponding lattice matching relation of the $CsPbBr_3$–$CsPb_2Br_5$ interface and Figure 3h,i show the lattice models of tetragonal $CsPb_2Br_5$ and monoclinic $CsPbBr_3$. The interface is composed of the parallel $CsPb_2Br_5$:(200) plane and $CsPbBr_3$:(110). The separation angle between the $CsPb_2Br_5$:(200) plane and $CsPbBr_3$:($\bar{2}$00) plane is 45°. The $CsPb_2Br_5$:(002) planes and $CsPbBr_3$:($\bar{1}$10) planes are perpendicular to the interface. We have d(002) = 7.63 Å, 2×d($\bar{1}$10) = 8.20 Å, so the lattice mismatch between them is 7.2%; the lattice mismatch between the $CsPb_2Br_5$:(200) plane (d = 4.25 Å) and $CsPbBr_3$:(110) plane (d = 4.13 Å) is 2.9%. In addition, Figure 3d,f also shows the lattice fringe and diffraction spots belonging to (004) lattice planes (d = 3.80 Å) of $CsPb_2Br_5$. As shown in the lattice model of Figure 3h,i, the $CsPb_2Br_5$:(004) planes (d = 3.81 Å) and $CsPbBr_3$:($\bar{1}$10) planes (d = 4.10 Å) are parallel to each other, and they constitute the upper interface shown in Figure 3d, with a lattice mismatch of 7.3%. The $CsPb_2Br_5$:(300) planes and CsPbBr$_3$:(110) planes are perpendicular to this interface. We have 3×d(300) = 8.49 Å, 2×d(110) = 8.26 Å, so the lattice mismatch between them is 2.7%. Therefore, the lattice mismatch along

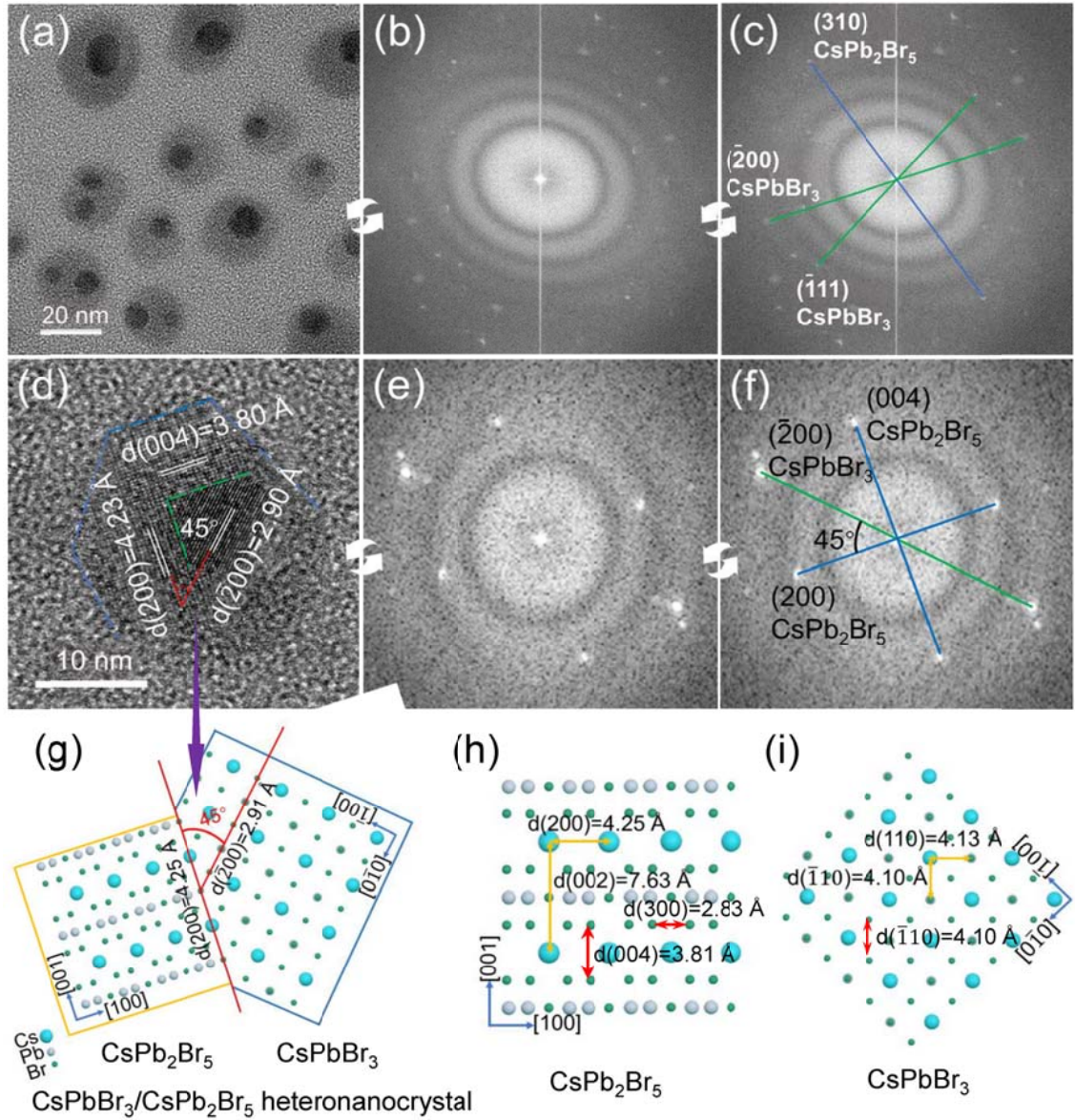

**Figure 3.** (a) HRTEM image of the CsPbBr$_3$/CsPb$_2$Br$_5$ core/shell HNCs in the white powder. (b, c) FFT of the image in (a). (d) HRTEM image of a typical HNC and (e, f) corresponding FFT.

Green and blue lines in (d) indicate core and shell regions, respectively. (g) Lattice matching relation at the interface of the $CsPbBr_3/CsPb_2Br_5$ core/shell HNC corresponding to that shown in (d). Crystal structures of (h) $CsPb_2Br_5$ and (i) $CsPbBr_3$ with marked lattice spacing values.

the direction either parallel or perpendicular to the interface is small, which ensures epitaxial crystallization of $CsPb_2Br_5$ and $CsPbBr_3$ at the interface of the HNC. The HRTEM observation reveals there is another type of epitaxial crystallization of the $CsPb_2Br_5$–$CsPbBr_3$ interface in the core/shell HNC, as shown in **Figure 4.** The FFT (Figure 4b,c) of the core/shell HNC shown in Figure 4a contains the bright spots belonging to the (200) and (220) planes of tetragonal $CsPb_2Br_5$, and the angle between them is measured to be 45°. There are some additional spots in the FFT image that belong to the ($\bar{2}01$) planes of $CsPbBr_3$. The magnified HRTEM image (Figure 4e) exhibits remarkable Moiré fringe in the core area, and it arises from the spatial overlap of the lattice planes of the $CsPbBr_3$ core and that of the $CsPb_2Br_5$ shell. The existence of the Moiré pattern demonstrates that the lattice planes of the core and those of the shell at the interface are nearly parallel and thus reveals the heteroepitaxial nature of the core–shell interface. The measurement of the lattice spacings indicates that the nearly parallel overlapped lattice planes are the (220) planes of $CsPb_2Br_5$ (d = 3.01 Å) and ($\bar{2}01$) planes of $CsPbBr_3$ (2.61 Å). This is consistent with the orientation relation of the corresponding FFT spots (Figure 4c). The lattice mismatch between these two groups of lattice planes is 13%, suggesting there can be strain at the interface. The separation angle between the $CsPb_2Br_5$:(220) planes and $CsPbBr_3$:($\bar{2}01$) planes is measured to be 8°, which agrees with the separation angle between the corresponding FFT spots (Figure 4c). The spacing of the Moiré fringe obeys $D = d_1 d_2 / \sqrt{(d_1 - d_2)^2 + d_1 d_2 \alpha^2}$, where $d_1$ and $d_2$ are the spacings of the nearly parallel and overlapped lattice planes of the core and shell, and $\alpha$ is the separation angle between them. We have $d_1$ = 3.01 Å [CsPb$_2$Br$_5$:(220) plane], $d_2$ = 2.61 Å [CsPbBr$_3$:($\bar{2}$01) plane], and $\alpha$ = 8°, so $D$ is calculated to be 1.4 nm, which is consistent with the measured value of 1.45 nm (Figure 4e). The HRTEM observation reveals that the CsPbBr$_3$ core and CsPb$_2$Br$_5$ shell have different ways of lattice matching at their epitaxial interface.

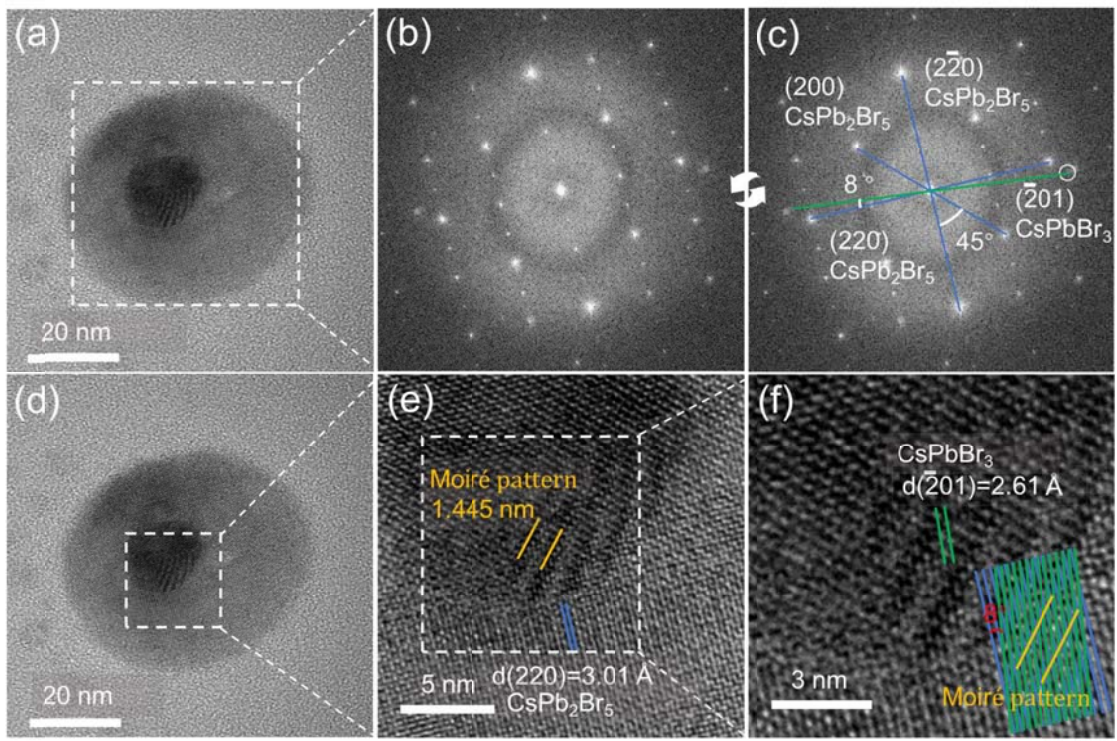

**Figure 4.** (a, d) HRTEM image of a typical CsPbBr$_3$/CsPb$_2$Br$_5$ core/shell HNC in the white powder. (b, c) Corresponding FFT. (e, f) Enlarged image of the marked area in (d).

**2.2. Optical Properties and Electronic Structures of CsPbBr$_3$/CsPb$_2$Br$_5$ HNCs.** The orange powder exhibited rather weak luminescence in that it contains micrometer-sized composites composed of agglomerated $CsPbBr_3$ NCs. In contrast, the white powder composed of $CsPbBr_3/CsPb_2Br_5$ core/shell HNCs emits bright green light. As shown in Figure S5a, the PL peak of the $CsPbBr_3$ NCs lies at 525 nm, which is consistent with the corresponding exciton absorption peak, indicting the exciton emission nature. The PL peak of the $CsPbBr_3/CsPb_2Br_5$ core/shell HNCs and that of the CsPbBr$_3$ NCs (Figure S5b) are very close, and this suggests that they have the same origin of luminescence, that is, arising from CsPbBr$_3$. It should be noted that the shell growth reduces the number of surface dangling bonds with potential trap states for the carriers and thereby improves the fluorescence intensity; however, it does not necessarily cause blue shift of the luminescence.[55] Figure 5a shows the PL, PL excitation (PLE), and UV–Vis absorption spectra of the white powder containing the CsPbBr$_3$/CsPb$_2$Br$_5$ HNCs. The PL spectrum is narrow, being characteristic of the exciton luminescence of the metal halide perovskites. Its maximum lies at 525 nm (2.36 eV). The PL peak remains nearly fixed when the excitation photon energy is changed (Figure S6) and this suggests there is only weak quantum confinement effect[56] in that the sizes of the core CsPbBr$_3$ NCs are bigger than the exciton Bohr diameter. The quantum yield of the CsPbBr$_3$/CsPb$_2$Br$_5$ HNCs was measured to be 13.0% under 374 nm excitation. In contrast, the quantum yield of the CsPbBr$_3$ NCs was as small as 0.07% (this value is not accurate enough because of the very weak luminescence signal). The UV–Vis absorption spectrum comprises two rising peaks at 517 nm (2.40 eV) and 343 nm (3.62 eV), respectively. The lower energy peak is sharp (this can be seen more clearly from the square of the absorbance versus photon energy curve shown in Figure 5b, although the exciton absorption does not follow the same law as band edge absorption), and it is the first exciton absorption peak of the core $CsPbBr_3$ NC. The PLE spectrum (emission: 550 nm) comprises also this exciton absorption peak. The higher energy peak at 3.62 eV is attributed to the exciton absorption of the $CsPb_2Br_5$ nanoshell, and this is consistent with the fact that $CsPb_2Br_5$ has a much bigger band gap compared with $CsPbBr_3$, as indicated by the following calculation. There are several additional even higher energy absorption peaks lying at 3.99, 4.58, and 4.84 eV, which are ascribed to the $CsPb_2Br_5$ nanoshell. From Figure 5a we notice a very interesting phenomenon: the rising of the UV–Vis band and decreasing of the PLE band in the range from 3.26 to 3.62 eV exhibit perfect mirror-symmetry correlation. This is because that starting from 3.26 eV the $CsPb_2Br_5$ nanoshell begins to absorb the incident photons more and more strongly, as a result, fewer and fewer photons can enter the inner $CsPbBr_3$ core, and thus the PL excitation intensity decreases in accordance. This perfect symmetric correlation between the shell-absorption increase and core-emission decrease confirms the core–shell structure of the $CsPbBr_3/CsPb_2Br_5$ HNCs. The PLE intensity decreases to nearly zero when the excitation photon energy exceeds 3.62 eV (Figure 5a) and this indicates that in this circumstance nearly no photons can penetrate the $CsPb_2Br_5$ shell to reach the $CsPbBr_3$ core and that there is no noticeable energy transfer from the shell to the core. This suggests that after the $CsPb_2Br_5$ shell absorbs higher-energy photons, the absorbed energies will be released to become thermal energies through efficient nonradiative recombination. From Figure 5b we see that the sharp exciton absorption peak of the $CsPbBr_3$ NCs shifts slightly to a lower energy of ~2.36 eV in the orange powder. This is because the average size of the $CsPbBr_3$ NC cores of the $CsPbBr_3/CsPb_2Br_5$ core/shell HNCs in the white powder is smaller than that of the original $CsPbBr_3$ NCs in the orange powder due to the phase transition of the outer layers of the latter into the $CsPb_2Br_5$ nanoshell. According to the quantum confinement effect, the smaller nanoparticle has a larger energy gap, therefore, the $CsPbBr_3$ core NCs in the HNCs have a higher energy exciton absorption peak compared with the $CsPbBr_3$ NCs in the orange powder. Additionally, the $CsPb_2Br_5$ nanoshell forms a potential energy barrier for the electrons and holes in the $CsPbBr_3$ core and this slightly changes the energy gap of the latter. The influence of the potential energy barrier on the energy gap of the encapsulated semiconductor nanocrystal is because the potential energy barrier is the critical quantity in the Schrödinger equation which determines the quantum states and energy levels of the crystals.[57] Figure S7 shows the calculated band alignment of monoclinic $CsPbBr_3$/tetragonal $CsPb_2Br_5$ heterostructure by using the Fermi level equivalence law of a semiconductor heterostructure in thermal equilibrium. The calculation shows that there is an energy barrier of 0.91/1.06 eV between the conduction band minima/valence band maxima of $CsPbBr_3$ and $CsPb_2Br_5$. Although the PL spectrum of the $CsPbBr_3$/$CsPb_2Br_5$ core/shell HNCs seems to have only one maximum, however, unlike a usual emission band that follows the Gaussian function as a result of dominant inhomogeneous broadening, it cannot be fitted by using a single Gaussian function, and neither can it be fitted by a single Lorentzian function corresponding to homogeneous broadening. Instead, it can be well fitted by using a double-Gaussian function, giving rise to two band peaks centered at 524.9 nm (2.36 eV) and 526.7 nm (2.35 eV), respectively. The former sub-band has a small full width at half maximum (FWHM) of 64.3 meV, and the latter has a much bigger FWHM of 130.9 meV. This suggests that there are two radiative recombination channels in the $CsPbBr_3$/$CsPb_2Br_5$ core/shell HNCs. Because the XRD patterns (Figure 2) show no signals of other types of NCs, so the two radiative recombination channels are hardly related to different types of NCs. To verify this conjecture, we measure the time-resolved PL spectra of the orange and white powders (Figure 5d). Both spectra can be fitted by using a biexponential function. The lifetimes (fractional intensities) of the two components for the $CsPbBr_3$ NCs are 8.30 μs (73.81%) and 384.3 ns (26.19%), respectively, and they decrease to 1.25 μs (78.86%) and 46.7 ns (21.14%) for the $CsPbBr_3/CsPb_2Br_5$ HNCs. These two decay channels indicate there are two radiative recombination channels in these HNCs which correspond to the two emission bands peaked at 524.9 and 526.7 nm. The decrease of the lifetimes of the two relaxation channels in the $CsPbBr_3/CsPb_2Br_5$ HNCs with respect to that in the pure $CsPbBr_3$ NCs may be caused by the influence of the potential barrier provided by the $CsPb_2Br_5$ nanoshell. The potential barrier offered by the $CsPb_2Br_5$ shell causes stronger spatial confinement of carriers, as a result, there could be higher degree of overlap of the wavefunctions of the electron and hole and thus the radiative quantum transition rate will be improved, this causes decrease of the lifetime. It should be noted that the reabsorption effect may also lead to two emission bands.[58]

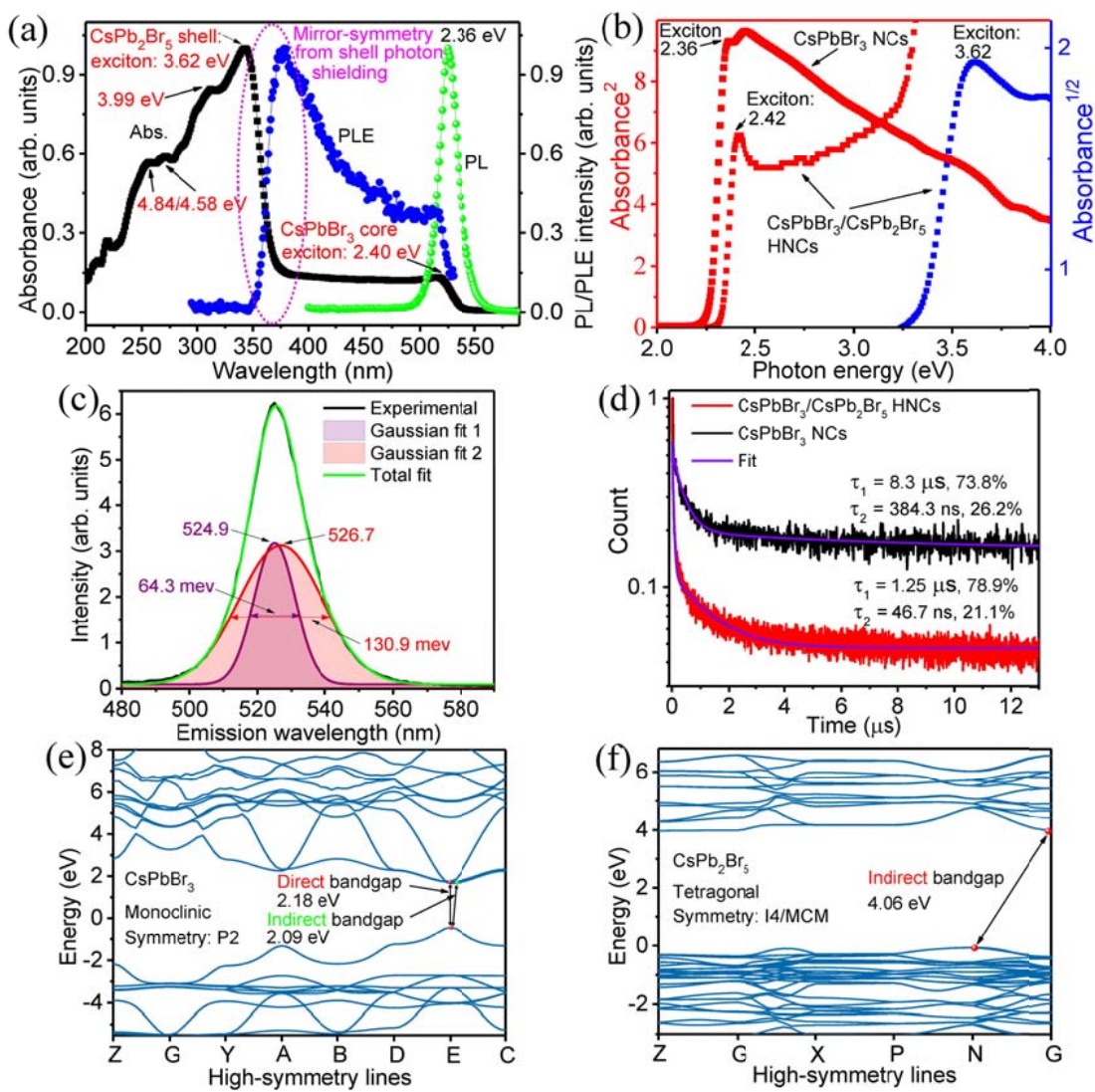

**Figure 5.** (a) UV–Vis absorption, PL (excitation: 380 nm), and PL excitation (emission: 550 nm) spectra of the $CsPbBr_3/CsPb_2Br_5$ core/shell HNC white powder. (b) Square and square root of absorbance versus photon energy for the CsPbBr$_3$/CsPb$_2$Br$_5$ HNCs and squared absorbance vs. photon energy for CsPbBr$_3$ NCs in the orange powder. (c) Double-Gaussian fit of the PL spectrum of the CsPbBr$_3$/CsPb$_2$Br$_5$ HNCs (excitation: 380 nm). (d) Time-resolved PL spectra and biexponential fits for the CsPbBr$_3$ NCs and CsPbBr$_3$/CsPb$_2$Br$_5$ HNCs (excitation: 370 nm; emission: 520 nm), respectively. Calculated electronic structure of (e) monoclinic CsPbBr$_3$ and (f) tetragonal CsPb$_2$Br$_5$ by using the DFT method employing the HSE06 exchange–correlation functional.

Because the optical properties are closely correlated with the electronic properties, therefore, we further calculate the electronic structures of the bulk monoclinic $CsPbBr_3$ and tetragonal $CsPb_2Br_5$ (Figure 5e,f) by using the density functional theory (DFT) calculation employing the HSE06 hybrid exchange–correlation functional. The monoclinic $CsPbBr_3$ has *P2* symmetry, and the original unit cell parameters employed are $a = b = 5.827$ Å, $c = 5.891$ Å ($\alpha = \beta = 90°$, $\gamma = 89.65°$) (PDF#18-0364). These parameters only deviate slightly from those of cubic $CsPbBr_3$ and thus the electronic and optical properties of monoclinic $CsPbBr_3$ may resemble those of the widely studied cubic $CsPbBr_3$. The calculation shows that monoclinic $CsPbBr_3$ has a direct band gap of 2.18 eV as well as an indirect band gap of 2.09 eV (Figure 5e), whose maximum lies at the proximity of the maximum of the direct gap, and their energy difference is only 90 meV. The existence of such an indirect band gap may be related to the Rashba effect. In some perovskites there is strong spin–orbit coupling due to the existence of the heavy elements (such as Pb and Br) which dominate the electron bands near the conduction band minimum and valence band maximum, it may result in large Rashba splitting if the crystal lacks inversion symmetry and in this circumstance the spin-degenerate parabolic band splits into two spin-polarized bands.[59–61] As a result, the perovskite has an indirect band gap which is slightly smaller than the direct band gap ($\Delta E \approx 50$ meV).[59,62] The DFT calculation shows that tetragonal $CsPb_2Br_5$ (*I4/MCM* symmetry) is an indirect band gap semiconductor with a big band gap of 4.06 eV (Figure 5f). The indirect gap nature explains why the $CsPb_2Br_5$ nanoshells in the $CsPbBr_3/CsPb_2Br_5$ HNCs do not emit light. This agrees with the previous observation that the $CsPb_2Br_5$ nanosheets have no efficient luminescence.[63] Note that the calculated band gap of $CsPb_2Br_5$ is bigger than the energy of the first exciton peak (3.62 eV) and close to that of the second absorption peak (3.99 eV) (Figure 5a), and this suggests that the second absorption peak may be related to the electron transition from the valence band to the conduction band, and it also implies that $CsPb_2Br_5$ has a big exciton binding energy, which is characteristic of the large band gap semiconductors.[11,12,64] As mentioned above, the DFT calculation reveals that monoclinic $CsPbBr_3$ has a direct band gap (2.18 eV) that is close to the indirect band gap (2.09 eV). In accordance, there may be simultaneously direct gap radiative recombination and indirect gap radiative recombination, the former is a faster decay, whereas the latter is a slow decay because the indirect-gap luminescence is an inefficient multi-order quantum process due to participation of the phonons in the electron recombination process.[12,65] Note that although the DFT calculations were performed for bulk crystals, the currently investigated nanocrystals with sizes of a few tens of nanometers have very weak quantum confinement effect and thus the DFT calculation results can well explain their properties.

The Raman spectroscopy gives more information about the microstructures and optical properties of the CsPbBr$_3$ NCs and CsPbBr$_3$/CsPb$_2$Br$_5$ HNCs. Figure 6a shows the photos of the orange and white powders under illumination of the halogen lamp. In the photo for the orange powder, the areas in brown color are CsPbBr$_3$ and the small white spots are CsPb$_2$Br$_5$. Figure

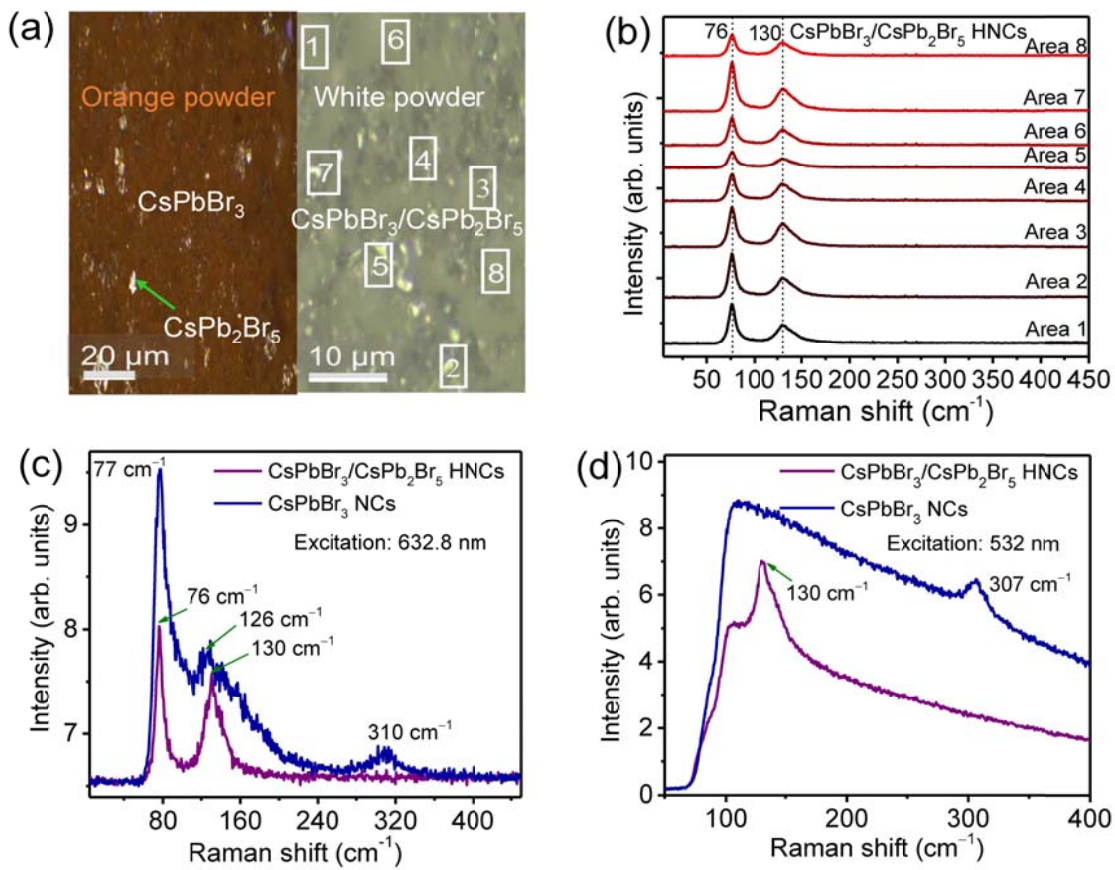

**Figure 6.** (a) Photographs of the CsPbBr$_3$ nanocrystal orange powder and the CsPbBr$_3$/CsPb$_2$Br$_5$ heteronanocrystal white powder under irradiation of the halogen lamp of the optical microscope. The white spots in the left part represent a small amount of CsPb$_2$Br$_5$ mixed in the orange powder. (b) Raman spectra of the CsPbBr$_3$/CsPb$_2$Br$_5$ heteronanocrystal white powder under 632.8-nm excitation for the eight marked regions in (a). Raman spectra of the CsPbBr$_3$/CsPb$_2$Br$_5$

heteronanocrystal white powder and the CsPbBr$_3$ nanocrystal orange powder measured under (c) 632.8-nm excitation and (d) 532-nm excitation, respectively.

6c,d show the Raman spectra of two powders measured under 632.8- and 532-nm excitation, respectively. The Raman peak at around 307–310 cm$^{-1}$ is characteristic of CsPbBr$_3$.[51,66] It is strong in the orange powder and absent in the white powder, suggesting that all the CsPbBr$_3$ NCs in the white powder should be completely encapsulated by the CsPb$_2$Br$_5$ nanoshells. The wide Raman peak at around 126 cm$^{-1}$ in the Raman spectrum of the orange powder is also characteristic of CsPbBr$_3$.[67] In contrast, the Raman spectra of the white powder comprise two sharp peaks centered at 76 and 130 cm$^{-1}$, which are mainly contributed by CsPb$_2$Br$_5$.[51,67] On the other hand, the energy dispersive X-ray spectroscopy (EDS) (Figure S8) of the scanning electron microscopy (SEM) indicates that the atomic ratio of Cs: Pb: Br equals 1: 1.7: 4.8 for the white powder (Table S2), which is close to the atomic ratio of the stoichiometric CsPb$_2$Br$_5$, suggesting that there are no exposed CsPbBr$_3$ nanoparticles in the white powder. This is confirmed by the Raman spectra measured at different areas of the white powder (Figure 6b), which exhibit no signal at 307–310 cm$^{-1}$ that is characteristic of CsPbBr$_3$. This result is expected in that the CsPbBr$_3$/CsPb$_2$Br$_5$ HNCs were generated in water via in situ phase transition of the original CsPbBr$_3$ NCs and any exposed CsPbBr$_3$ NCs decompose in water. The EDS (Figure S9) reveals that the atomic ratio for the orange powder is Cs: Pb: Br = 1.2: 1.0: 3.3 (Table S2), being consistent with that of stoichiometric CsPbBr$_3$ within the experimental error. This supports that the dominant component in the orange powder is CsPbBr$_3$. Note that the asymmetric broad Raman peaks (Figure 6) characteristic of CsPbBr$_3$ and CsPb$_2$Br$_5$ suggest there could be several very close vibrational modes which superimpose to give rise to one seemingly asymmetric and broad peak.

We further study the thermal stability of the $CsPbBr_3$/$CsPb_2Br_5$ HNCs by measuring the temperature dependent PL spectra of the $CsPbBr_3$/$CsPb_2Br_5$ powder starting from 78 K (Figure 7a). When the temperature increases, the PL spectrum shifts to blue monotonically from 2.362 to 2.380 eV (Figure 7b). The absence of abrupt PL change (peak wavelength and linewidth) suggests there is no phase transition in this temperature range. The increase of energy gap with temperature is characteristic of the metal halide perovskites[64] and this is contrary to that of the conventional semiconductors.[68–70] This abnormal temperature dependence was supposed to arise from the specific thermal expansion and electron–phonon coupling of the metal halide perovskites.[71] The linewidth of the PL spectrum increases from 0.059 to 0.106 eV as the temperature increases from 78 to 293 K (Figure 7b) and this may be caused by the enhanced electron–phonon coupling at the elevated temperatures.[64,71,72] Furthermore, the PL intensity drops gradually with increasing temperature (Figure S10) because of involvement of more phonons and enhanced nonradiative recombination.[73] Note that the crystal structure of the $CsPbBr_3$/$CsPb_2Br_5$ HNC powder remains unchanged during increase of temperature, as indicated by the XRD measurement (Figure S11).

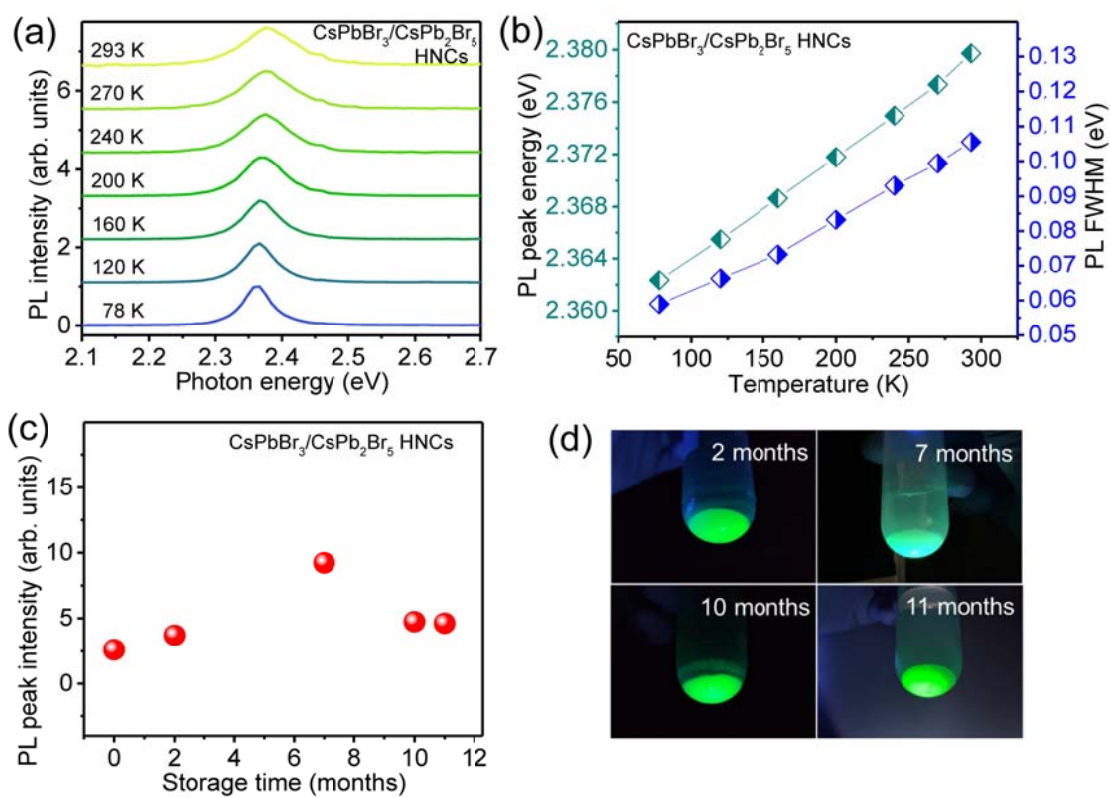

**Figure 7.** (a) PL spectra (excitation: 370 nm) of the $CsPbBr_3$/$CsPb_2Br_5$ HNC powder measured at different temperatures. (b) Plot of PL peak energy and linewidth versus temperature. (c) PL peak intensity versus storage time for the $CsPbBr_3$/$CsPb_2Br_5$ HNC powder immersed in water. (d) Corresponding luminescence photos taken under 365-nm excitation.

As aforementioned, although the perovskites have excellent optical properties, they are usually very unstable upon exposure to heat or water. Among these unfavorable factors, water erosion to the perovskites is the most lethal. Many types of perovskites decompose rapidly in water and this causes significant decrease of the luminescence intensity. Surprisingly, we find that after the $CsPbBr_3$/$CsPb_2Br_5$ HNC powder has been immersed in water for over eleven months, its luminescence intensity remains nearly unchanged (Figure 7c,d). Figures S12 and S13 show the XRD pattern as well as PL and PLE spectra of the $CsPbBr_3$/$CsPb_2Br_5$ HNC powder after immersion in water for 11 months. Comparison with the characterizations of the pristine powder reveals both the crystal structure and the luminescence properties remain unchanged after the storage (the slight broadening of the XRD peaks for the aged sample may be caused by aggregation of some smaller nanoparticles). This proves that the $CsPbBr_3$/$CsPb_2Br_5$ HNCs have excellent water resistance and thus the $CsPb_2Br_5$ nanoshells can effectively protect the green luminescent $CsPbBr_3$ NC cores from being eroded by water. By contrast, as shown in Figure S14, the luminescence of the pure $CsPbBr_3$ quantum dots suspended in toluene (see EXPERIMENTAL section for the synthesis method) vanished completely after water was added into the solution followed by shaking for 5 minutes. We further studied the thermal stability of the $CsPbBr_3/CsPb_2Br_5$ HNCs. The $CsPbBr_3/CsPb_2Br_5$ HNCs placed in a plate was heated to a specific high temperature and maintained for 0.5 h, then the PL spectrum was measured immediately right after the powder cooled down. It is found that the powder can retain 81.3% of the original luminescence intensity (at room temperature) for a test temperature of 100 °C, and for even higher test temperatures of 150 and 200 °C, the $CsPbBr_3/CsPb_2Br_5$ HNCs can still retain 46.7% and 19.7% of its original luminescence intensity (Figure S15). This excellent thermal stability is very impressive and suggests that the luminescent $CsPbBr_3/CsPb_2Br_5$ HNCs can work at high temperatures. This virtue renders them superior solid-state light emitters compared with the perovskites protected by other types of shells.[74–76]

**2.3. Quasi-White LED Based on $CsPbBr_3/CsPb_2Br_5$ Core/Shell HNCs.** Because the synthesized luminescent $CsPbBr_3/CsPb_2Br_5$ HNCs have a large yield and because they have excellent water and thermal stability, hence they may serve as the backlights in the displays. We fabricated a prototype quasi-white LED by spin coating the mixture of the $CsPbBr_3/CsPb_2Br_5$ HNC powder and red luminescent phosphor $K_2SiF_6:Mn^{4+}$ (KSF) on the violet-emitting (380 nm) GaN chip. Figure 8b shows the electroluminescence (EL) spectrum of this LED operating at 3 V and 10 mA, and apparently it comprises three primary colors. Figure 8c displays the corresponding electroluminescence photo of the LED. It can be seen that this device emits bright quasi-white light. Figure 8a shows the Commission Internationale de L'Eclairage (CIE) chromaticity diagram of this LED. Its color coordinates are (0.4024, 0.4647), and its color gamut is 1.2 times that of the National Television System Committee (NTSC) standard. The color coordinates of the green light from the $CsPbBr_3/CsPb_2Br_5$ HNC powder are (0.1853, 0.7530), while that of the NTSC standard green light are (0.21, 0.71) (Figure 8a and Table S3), hence the green light emitted by the $CsPbBr_3/CsPb_2Br_5$ HNC powder has a wider color gamut. We have also conducted the stability test of the LED device. Figure S16 shows the normalized relative intensity ratio of $I_{521\ nm}$ ($CsPbBr_3/CsPb_2Br_5$ HNCs): $I_{380\ nm}$ (GaN chip)) versus operation time for the LED based on the core–shell HNCs. It can be seen that after 480 min of continuous operation, the luminescence intensity of the $CsPbBr_3/CsPb_2Br_5$ HNC emitting layer with respect to that of the GaN chip drops to a still high value of 60.8%. This indicates that the $CsPbBr_3/CsPb_2Br_5$ HNCs have excellent structural and luminescence stability. These characteristics suggest they are strongly competent materials for display backlights.

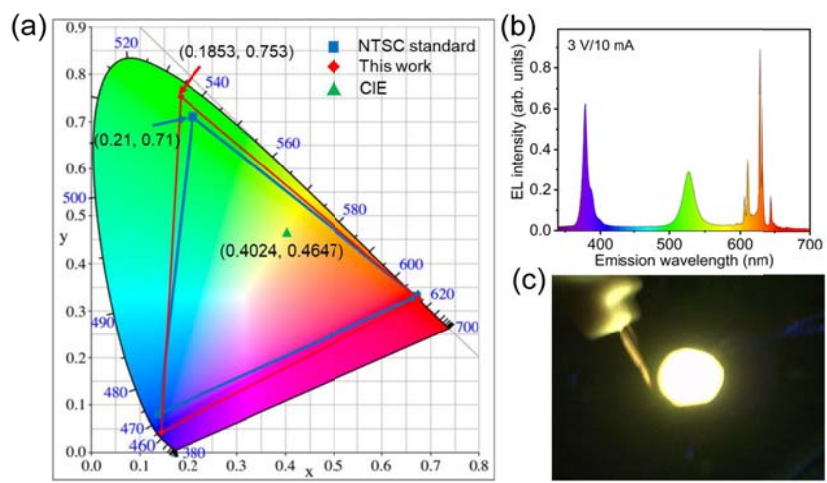

**Figure 8.** Characterization of a quasi-white LED constructed from a GaN chip coated by the mixture of the green luminescent $CsPbBr_3/CsPb_2Br_5$ HNC powder and red luminescent KSF phosphor. (a) CIE chromaticity coordinates (0.4024, 0.4647) of the LED along with its color triangle diagram and the NTSC standard triangle diagram. (b) Electroluminescence spectrum and (c) luminescence photo of this LED operating at 3 V and 10 mA.

## 3. CONCLUSIONS

In summary, we have reported the high-yield synthesis of the highly water-resistant green luminescent $CsPbBr_3/CsPb_2Br_5$ core/shell HNCs and deeply investigated their optical properties. The water-driven in situ self-phase transition of the original monoclinic $CsPbBr_3$ NCs leads to formation of the epitaxial tetragonal $CsPb_2Br_5$ nanoshells surrounding the luminescent $CsPbBr_3$ cores. The microstructural characterizations reveal that the small lattice mismatch between $CsPbBr_3$ and $CsPb_2Br_5$ accounts for the epitaxial crystallization at their interface. The optical characterizations in conjunction with the hybrid DFT calculations indicate that tetragonal $CsPb_2Br_5$ is an indirect wide band gap semiconductor and the $CsPb_2Br_5$ nanoshell forms an electron potential barrier surrounding the $CsPbBr_3$ core, which causes the largely enhanced double-channel exciton luminescence of the $CsPbBr_3$ core. These waterproof thermally-stable purely-inorganic perovskite semiconductor–semiconductor core–shell heteronanocrystals are very competent in solid-state lighting and wide color gamut displays.

## 4. EXPERIMENTAL AND COMPUTATIONAL DETAILS

**4.1. Materials.** $PbBr_2$ (99%, Aladdin), CsBr (99.9%, Aladdin), hydrobromic acid (HBr, 48%, Chron), and $K_2SiF_6:Mn^{4+}$ (KSF, Looking Long). All chemical materials were used directly without further purification.

**4.2. Synthesis of Monoclinic CsPbBr$_3$ Nanocrystal Powder.** PbBr$_2$ (10 mmol) were dissolved in HBr (8 mL), and CsBr (10 mmol) was dissolved in deionized water (3 mL). Then the CsBr solution was added dropwise to the PbBr$_2$ solution in an ice–water bath, and the mixed solution was stirred for 1 h, during which the orange precipitate containing mainly the monoclinic CsPbBr$_3$ NCs was produced rapidly. The precipitate was filtered and washed three times with ethanol and then dried in a vacuum drying oven at 60 °C in the dark for 12 h.

**4.3. Synthesis of CsPbBr$_3$/CsPb$_2$Br$_5$ Heteronanocrystal Powder.** The deionized water was added into the as-prepared CsPbBr$_3$ NC powder, during which the color of the powder changed gradually from orange to white. After stirring for 1.5 h, the powder turned into white completely, which was collected by using the suction filtration. Subsequently, it was dried in a vacuum drying oven at 60 °C in the dark for 12 h.

**4.4. Synthesis of CsPbBr$_3$ quantum dots.** The CsPbBr$_3$ quantum dots with a most probable size of 6.8 nm were synthesized by using the conventional method.[4] PbBr$_2$ (0.069 g) and 1-octadecene (ODE) (5 mL) were added into a three-necked flask and then dried at 120 ºC for 2 h under an argon atmosphere. Then, the dry oleylamine (OLA) (0.5 mL) and oleic acid (OA) (0.5 mL) were injected into the flask at 120 °C under an argon atmosphere. After the PbBr$_2$ salt was completely dissolved, the reaction temperature was improved to 140 °C, and then Cs-oleate solution (0.4 mL) was quickly injected into the three-necked flask. After 5 s, the reaction solution was cooled to room temperature in an ice–water bath. After addition of tert-butanol (tBuOH, ODE:tBuOH = 1:1 in volume ratio), the crude solution was centrifuged to generate precipitated CsPbBr$_3$ quantum dots, which were then dissolved in toluene.

**4.5. Fabrication of the Quasi-White LED.** 0.36 g CsPbBr$_3$/CsPb$_2$Br$_5$ powder and 0.18 g red luminescent phosphor K$_2$SiF$_6$:Mn$^{4+}$ (KSF) powder were introduced into 1.217 g organic silica gel, and the mixture was stirred evenly with a glass rod. The mixed glue was uniformly coated onto a GaN chip (emission: 380 nm). It was then put in a blast drying box and annealed at 100 °C for 1 h.

**4.6. Characterization.** The Fluorlog3-TCSPC spectrofluorometer (HORIBA JOBIN YVON) equipped the Xe lamp (excitation wavelengths employed: 360–500 nm) and pulsed laser diodes was used to measure the PL, PL excitation, and time-resolved PL spectra at room temperature. The quantum yield was acquired by using the integration sphere under excitation of a Xe lamp. The low temperature PL spectra were measured under 370-nm excitation by cooling the samples with liquid nitrogen. The TEM characterization was performed on a Tecnai G2 T20 TEM (FEI) operating at 200 kV. The Smartlab (3) intelligent X-ray diffractometer was used for the X-ray diffraction characterization. The UV–Vis absorption spectra were measured by using a UV–Vis spectrophotometer (HITACHI UV-3600Plus). The LabRAM HR 800 Raman spectrometer (HORIBA JOBIN YVON) and Witec Alpha300RA Raman spectrometer were used to conduct the Raman spectroscopy under 532-nm and 632.87-nm laser excitation. The scanning electron microscope (FEI Inspect F50) equipped with an energy dispersive X-ray spectrometer (EDS) probe was used for the SEM observation and elemental analysis. The Acton SP-2358 Spectrometer (Princeton Instruments) was used to measure the electroluminescence spectra. The bias was applied on the LED by using a Keithley 2400 source meter in air at room temperature.

**4.7. Computation Methods.** The density functional theory calculations were implemented in the Vienna ab initio simulation package (VASP) code.[77] The interaction between the ions and electrons was described by using the projector augmented wave (PAW) method.[78] The Perdew–Burke–Ernzerhof (PBE) exchange–correlation functional[79] within the generalized gradient approximation (GGA) was employed for the geometric optimization with a force tolerance for ionic relaxation of 0.3 eV/Å. The HSE06 hybrid functional[80,81] was used for the calculation of the electronic structure. The cut-off energy of 300 eV (accurate enough, as can be seen from the convergence test shown in Figure S17) was used for the plane-wave expansions in all the calculations. The calculations were performed using a 4×4×4 Gamma centered grid. The band alignment of monoclinic $CsPbBr_3$/tetragonal $CsPb_2Br_5$ heterostructure (Figure S7) was calculated by using the Fermi level equivalence law of a semiconductor heterostructure in thermal equilibrium.[80] The Fermi level of an intrinsic semiconductor follows:[82]

$$E_{Fi} = E_v + \frac{1}{2}E_g - \frac{1}{2}k_B T \ln(\frac{N_c}{N_v}), \qquad (1)$$

where $k_B$ is the Boltzmann constant, $T$ (300 K) is the absolute temperature, $E_v$ and $E_g$ are the valence band maximum energy level and bandgap, and the effective density of states at the conduction/valance band edge is given by:

$$N_c = 2[2\pi m_e^* k_B T/h^2]^{3/2} \qquad (2)$$

$$N_v = 2[2\pi m_h^* k_B T/h^2]^{3/2} \qquad (3)$$

In these equations, $m_e^*$ and $m_h^*$ are the effective mass of electron at the conduction band minimum and that of hole at the valence band maximum, h is the Planck constant. The effective mass-tensor was calculated by using the effective mass calculator (EMC) code[83] that is based on the VASP calculation results. The calculated bandgaps, effective masses, and Fermi level shift relative to the middle point of the forbidden gap are listed in Table S1.

## ACKNOWLEDGMENT

This work was supported by the National Natural Science Foundation of China No. 11874106.

Perovskite Nanocrystals Incorporated into Polystyrene Microspheres for Long-Term Storage and Reusage. *Angew. Chem. Int. Ed.* **2019**, *58*, 2799.

(77) Kresse, G.; Furthmüller, J. Efficient Iterative Schemes for ab initio Total-Energy Calculations Using a Plane-Wave Basis Set. *Phys. Rev. B* **1996**, *54*, 11169.

(78) Kresse, G.; Joubert, D. P. From Ultrasoft Pseudopotentials to the Projector Augmented-Wave Method. *Phys. Rev. B* **1999**, *59*, 1758.

(79) Perdew, J. P.; Burke, K.; Ernzerhof, M. Generalized Gradient Approximation Made Simple. *Phys. Rev. Lett.* **1996**, *77*, 3865–3868.

(80) Heyd, J.; Scuseria, G. E.; Ernzerhof, M. Hybrid Functionals Based on a Screened Coulomb Potential. *J. Chem. Phys.* **2003**, *118*, 8207–8215.

(81) Krukau, A. V.; Vydrov, O. A.; Izmaylov, A. F.; Scuseria, G. E. Hybrid Functionals Based on a Screened Coulomb Potential. *J. Chem. Phys.* **2006**, *125*, 224106.

(82) Kasap, S. O. *Optoelectronics and Photonics: Principles and Practices*, 2nd ed.; Pearson Education, Inc.: Upper Saddle River, 2013,

(83) Fonari, A.; Sutton, C. Effective Mass Calculator, 2012.